\documentclass[preprint,5p, twocolumn, compress]{elsarticle}

\usepackage{amssymb}
\usepackage{xurl}
\usepackage[hidelinks]{hyperref}
\usepackage{amsmath}
\usepackage{latexsym}
\usepackage{enumitem}
\usepackage{graphicx} 
\usepackage{color}
\usepackage{times}
\usepackage{soul}
\usepackage{multirow}
\usepackage{amsthm}
\usepackage{booktabs}
\usepackage[switch]{lineno}
\usepackage{CJKutf8}
\usepackage{bbding}
\usepackage{amssymb}
\usepackage{tabularx}

\usepackage{titlesec}

\journal{Measurement}
\begin{document}
\begin{frontmatter}

\title{Towards a Bathroom-Centered Human–Building Digital Twin Framework for Indoor Safety Analysis} 

\author[1]{Yuanzhi Su}
\ead{yuanzhi.su@connect.polyu.hk}
\author[1]{Huiying (Cynthia) Hou\corref{corresponding}}
\ead{cynthia.hou@polyu.edu.hk}

\cortext[corresponding]{Corresponding author.}

\affiliation[1]{organization={Department of Building Environment and Energy Engineering},
                addressline={The Hong Kong Polytechnic University},
                city={Hong Kong},
                country={China}}

\begin{abstract}
Bathroom use is a critical safety challenge for older adults because wet surfaces, constrained layouts, limited support, and frequent posture transitions are concentrated within a small domestic space. These conditions create risks that cannot be adequately understood by considering either the bathroom environment or human motion in isolation. Existing bathroom safety studies mainly identify hazards, accessibility problems, or design modifications, whereas human-centered sensing studies often focus on activity recognition or fall detection without sufficient semantic understanding of the surrounding environment. This separation limits the interpretation of how older adults interact with fixtures, support surfaces, wet areas, and spatial constraints during daily bathroom activities. To address this gap, this study proposes a bathroom-centered human-building digital twin framework for interaction-aware indoor safety analysis with a specific emphasis on older adult bathroom safety. The framework conceptualizes bathroom risk as a coupled human-environment process and integrates semantic bathroom representation, skeleton-based human representation, spatial-semantic coupling, interaction-aware event analytics, and safety-oriented visualization. A Unity-based proof-of-concept prototype is developed to demonstrate the feasibility of the framework. Although the current work remains a prototype-oriented investigation, it establishes a methodological basis for analyzing older adults' bathroom safety through explicit body-environment relations and for advancing privacy-sensitive, interaction-aware digital twin applications in aging-in-place residential environments.
\end{abstract}


\end{frontmatter}

\section{Introduction}
Falls and injury events among older adults are a major public-health and residential-safety concern~\cite{kamei2015effectiveness, voljvc2016public}. Age-related changes in balance control, muscle strength, gait stability, vision, reaction time, and cognitive function can reduce an older adult's ability to recover from slips, trips, unstable transfers, or unexpected contact with surrounding objects~\cite{wang2024age, ccekok2025effects, hollman2007age, who2021falls}. However, these individual vulnerabilities do not operate in isolation. A central argument in environmental gerontology is that older adults' functional performance depends on the fit between personal competence and environmental press, meaning that the same physical setting may impose very different levels of difficulty on users with different mobility, sensory, and balance capacities~\cite{lawton1973ecology, iwarsson2005long, wahl2009home, lawton1999environmental}. From this perspective, older adult safety is not only a medical or behavioral issue, but also a person-environment problem shaped by housing design, accessibility barriers, assistive features, and the spatial organization of everyday activities~\cite{iwarsson1999housing,granbom2016public,lien2016assessment}. 

Bathrooms represent one of the most critical residential micro-environments for older adult safety~\cite{myers2018most, penner2013bathroom, mesthrige2020critical, gleisner2022towards}. Unlike many other domestic spaces, bathroom use combines water exposure, compact layouts, closely spaced fixtures, restricted turning space, and repeated posture transitions such as sitting, standing, bending, reaching, and transferring~\cite{jaglarz2020ergonomic, camara2010analysis, edwards2019scoping, rao2019ergonomic}. These conditions can create a strong environmental press on older users, particularly when reduced balance, slower gait, weaker lower-limb function, visual decline, or the use of mobility aids limits their ability to adapt to sudden slips, narrow clearances, or unstable transfers. Epidemiological evidence supports this concern: bathroom injuries are predominantly fall-related, injury rates increase with age~\cite{schellenberg2019falls, kim2023understanding}, and the highest-risk locations and activities are concentrated around the tub, shower, and toilet, especially during bathing, showering, getting out of the tub or shower, and toilet-related sit-to-stand activities~\cite{stevens2011nonfatal}. Therefore, bathroom safety for older adults should be understood as a coupled problem involving person-specific vulnerability, activity demands, and spatial-environmental conditions rather than as a simple collection of isolated hazards.

Gerontology, housing, and built-environment studies further show that bathroom risk is closely related to the everyday activities that allow older adults to maintain independence at home~\cite{percival2002domestic, brim2021older, gleisner2022towards}. Bathing and toileting are basic activities of daily living, yet they involve complex movements that are biomechanically and spatially demanding, including stepping across a wet boundary, rotating in a confined area, lowering to or rising from a toilet seat, reaching for towels or fixtures, and maintaining balance while washing or dressing. The risk profile of these activities is therefore affected not only by surface wetness, but also by the location and usability of grab bars, adequacy of slip-resistant surfaces, toilet and fixture height, shower or bathtub configuration, threshold conditions, lighting, ventilation, clutter, and reachable support surfaces~\cite{afifi2015geriatric,blanchet2018need,korp2012bathing,capezuti2008bed,ng2022bathroom,afacan2024fallfree}. Recent user-centered bathroom research also suggests that older adults' perceptions of a fall-free bathroom are associated with both supporting facilities and indoor environmental qualities, such as artificial lighting, ventilation, accessible towel rails, comfort, ease of access, error-proof design, and emergency management~\cite{afacan2024exploring}. These findings indicate that bathroom risk is not merely a matter of detecting a fall after it occurs; it is produced through the dynamic interaction between an older person's functional capacity, bathroom layout, fixture design, and activity-specific movement.

Existing research has examined older adult bathroom safety and residential fall prevention from several relevant but fragmented perspectives. Built-environment and gerontological studies have emphasized accessibility, environmental hazard identification, fixture arrangement, grab-bar provision, bathroom modification, and aging-in-place interventions~\cite{afifi2015geriatric,blanchet2018need,valipoor2020falls,ng2022bathroom,afacan2024fallfree}. This work is valuable because it establishes the bathroom as a safety-sensitive space and identifies design-related risk sources that are particularly relevant to older users. However, much of this literature remains descriptive, checklist-based, design-oriented, or intervention-oriented. It can indicate that a wet floor, missing support, narrow clearance, unsuitable toilet height, or poorly positioned fixture may increase risk, but it usually does not describe how an older adult's body actually moves through the space, how a transfer or turning movement unfolds, which object becomes involved, or where contact, instability, or impact occurs during a specific bathroom activity.

In parallel, human-centered sensing studies have advanced activity recognition, fall detection, and behavior monitoring for older adults and indoor safety applications using wearable sensors, ambient sensors, radar sensing, and vision-based approaches~\cite{delahoz2014survey,wang2020elderly,alam2022vision,jin2020mmfall,su2025posegraphnet}. Skeleton-based motion representation is especially relevant because it provides a compact and interpretable abstraction of posture, trajectory, and kinematic change while reducing reliance on raw video streams in privacy-sensitive residential spaces~\cite{sykes2025next,luo2022pervasive}. Nevertheless, many sensing systems remain person-centered rather than person-in-environment-centered. They often classify whether a fall or activity has occurred, but they provide limited information about the surrounding bathroom context: which fixture was nearby, whether the movement occurred in the shower or toilet zone, which body part contacted the environment, whether the event involved a support surface, or how spatial constraints shaped the event. For older adult bathroom safety, this limitation is substantial because fall-related events are often produced by the interaction between age-related vulnerability, activity-specific movement, and local environmental conditions.

This separation between environmental representation and human-motion analysis limits the interpretability of bathroom-centered safety assessment. A motion-centric output such as ``fall detected'' is insufficient for explaining why a safety-relevant event occurred, whether it involved a wet shower area, an unsupported toilet transfer, a narrow turning space, a level change, or contact with a nearby fixture. Conversely, a static room-centric model can document environmental risk factors but cannot explain how an older adult interacts with those risk factors during actual use. A more useful analytical approach should therefore represent the bathroom as a semantic environment, the older occupant as a dynamic human body, and the safety event as a body-environment interaction process unfolding over time. Such an approach is necessary if bathroom safety research is to move beyond general hazard description or isolated fall detection toward spatially grounded interpretation of older adults' daily bathroom use.

Digital twin (DT) research provides a promising basis for this integration. In the built environment, DTs have increasingly been used for spatial representation, monitoring, simulation, operational analysis, and decision support. Their value lies not merely in visualization, but in connecting physical entities, sensed data, computational models, and analytical results within a unified and updatable digital environment~\cite{albalkhy2024digital,mousavi2024digital,zhang2025review,backlund2024showcasing}. However, many existing building-oriented DT studies remain predominantly building-facing. They often focus on geometry, assets, environmental variables, energy performance, or facility operation, while occupant behavior is weakly represented or treated as an external condition. Related work on human-building interaction has emphasized the coupled relationship between occupants and spaces~\cite{taherkhani2023human,kim2023human}, but this perspective has not yet been sufficiently operationalized in DT frameworks for event-level indoor safety analysis, particularly in confined, high-risk residential spaces used by older adults.

For older adult bathroom safety, a bathroom-centered human-building DT should support at least three capabilities. First, it should represent the bathroom not as a generic room, but as a semantic environment containing safety-relevant objects, functional zones, support surfaces, and potential hazard regions. Second, it should represent the occupant dynamically through posture, motion state, trajectory, and kinematic descriptors that are relevant to balance loss, transfer difficulty, support-seeking behavior, and impact. Third, it should explicitly couple the human representation and the bathroom representation in a shared coordinate and semantic space so that body-environment relations can be analyzed over time. This shift is important because the safety of older adults in bathrooms depends not only on whether a hazardous environment exists or whether a fall is detected, but on how the older user's movement and the bathroom environment jointly produce risk.

In response to this need, this study proposes a bathroom-centered human-building digital twin framework for interaction-aware indoor safety analysis with a specific emphasis on older adult bathroom safety. The proposed framework is organized as a six-layer architecture: the physical sensing layer, data communication and management layer, human-building twin construction layer, spatial-semantic coupling layer, interaction-aware analytics layer, and visualization and decision-support layer. The physical sensing layer defines the real bathroom environment, sensing devices, and safety-relevant physical entities. The data communication and management layer supports the ingestion, synchronization, and organization of sensing and spatial data. The human-building twin construction layer builds a semantic bathroom twin and a skeleton-based human twin. The spatial-semantic coupling layer aligns the human twin with the bathroom coordinate system and maps body motion to semantic objects and functional zones. The interaction-aware analytics layer extracts behavior states, kinematic descriptors, contact-related events, and contextual risk indicators. Finally, the visualization and decision-support layer presents these analytical results through a DT interface for safety-oriented interpretation.

The present study should be understood as a prototype-oriented investigation. Its objective is to formalize older adult bathroom safety as a coupled human-environment process and to operationalize this perspective through a proof-of-concept DT prototype. The prototype demonstrates how human motion can be interpreted relative to bathroom semantics and transformed into interaction-aware descriptors that are relevant to older adult safety, including motion state, body-object proximity, contact events, impact-related information, and spatially contextualized risk representation. Hence, the main contributions of this study are as follows:
\begin{itemize}
    \item It proposes a bathroom-centered human-building DT framework that reframes older adult bathroom safety as a coupled human-environment problem involving user vulnerability, activity demands, and spatial-environmental risk sources.
    \item It develops a six-layer architecture that combines sensing-informed bathroom representation, skeleton-based human representation, and spatial-semantic coupling into a unified human-building DT workflow.
    \item It demonstrates a proof-of-concept prototype that supports semantic bathroom representation, occupant visualization, behavior-state understanding, body-environment event descriptors, and contextual risk representation for older adult indoor safety analysis.
    \item It provides a methodological foundation for future research on privacy-sensitive bathroom sensing, richer body-environment interaction analytics, human injury assessment, and broader human-building DT applications for aging-in-place and residential safety.
\end{itemize}

The remainder of this paper is organized as follows. Section~2 reviews relevant literature on built-environment DTs, human-centered digital representations, human-building interaction, and older adult bathroom safety. Section~3 describes the proposed six-layer DT architecture and its major components. Section~4 discusses the research significance, advantages, implications, and limitations of the current prototype. Section~5 concludes the paper and outlines future research directions.

\section{Literature review}
\subsection{Digital twins in the built environment}
Digital twin (DT) research in the built environment has expanded rapidly in recent years, but the field is still developing in terms of definition, architecture, and application boundaries~\cite{shahzad2022digital,albalkhy2024digital,abdelrahman2025digital}. Recent reviews show that building-oriented DT studies have mainly concentrated on lifecycle management, energy optimization, facility operation, maintenance, and environmental monitoring~\cite{albalkhy2024digital,mousavi2024digital,zhang2025review}, while implementation in the built-environment internal sector remains at a relatively early stage. These studies have clarified that a DT in buildings is not merely a static 3D model, but a data-connected representation that integrates geometry, sensing, analytics, and operational feedback~\cite{fuller2020digital,albalkhy2024digital,zhang2025review}. At the same time, the dominant emphasis of current building DT work remains on asset performance and system-level operation rather than occupant-centered safety analysis.~\cite{albalkhy2024digital, mousavi2024digital, zhang2025review, backlund2024showcasing} A related body of work has emerged under the concept of human–building interaction (HBI), which frames buildings not only as physical systems but also as spaces that shape human behavior, comfort, emotion, and activity~\cite{nembrini2017human,kim2023human,taherkhani2023human}. Bibliometric evidence indicates that HBI research has grown steadily, with a marked increase after 2016, reflecting broader interest in integrating human-centered computing and architectural thinking~\cite{taherkhani2023human}. However, much of this literature addresses comfort, usability, sustainability, and interactive environments at a broad conceptual level. It provides an important theoretical basis for linking occupants and buildings, but it does not by itself offer a concrete digital twin framework for event-level safety analysis in confined high-risk spaces such as bathrooms.

For the present study, these literature are highly relevant but not sufficient. Existing building DT research establishes the technological foundation for semantic spatial representation and data integration, while HBI research highlights the importance of studying the coupled relationship between occupants and spaces. Yet neither stream has adequately addressed how a bathroom-centered DT can represent behavior, object-level interaction, and risk-relevant events within a unified analytical framework.~\cite{albalkhy2024digital, taherkhani2023human}

\subsection{Human digital twins within indoor environments}
In parallel with the development of building DTs, human-centered digital representations have advanced through research on human modeling, pose estimation, activity recognition, fall detection, and behavior monitoring~\cite{he2024digital,su2025posegraphnet,aggarwal2014human,su2025high,noor2023lightweight,newaz2023methods}. In indoor safety applications, these studies commonly aim to infer human activities or abnormal events from wearable sensors, ambient sensors, radar signals, or vision-based data. Among these approaches, skeleton-based representation has become increasingly important because it provides a compact and interpretable description of human posture and motion. By representing the body as a set of joints and body segments, skeleton-based methods can support motion-state recognition, kinematic feature extraction, and fall-related event analysis while reducing the dependence on raw video streams in privacy-sensitive indoor environments~\cite{sykes2025next,luo2022pervasive,shi2019two,duan2022revisiting}.

However, most existing human motion monitoring systems remain person-centered rather than person-in-space-centered. For instance, they are often designed to classify whether a fall has occurred, distinguish abnormal events from normal activities, or recognize predefined activity categories. Although such methods are useful for event detection, they usually do not explicitly represent where the movement occurs, which nearby object or surface is involved, or how the body interacts with the surrounding environment during instability, support-seeking, or impact. In these systems, the indoor environment is often treated as background context rather than as a semantically structured and analytically active component of event interpretation. This limitation is particularly important for bathroom safety, where spatial confinement, fixture proximity, slippery regions, and support availability strongly influence how safety-relevant events unfold~\cite{shishov2024interactions,valipoor2020falls}. Accordingly, a skeleton-based human twin becomes more meaningful when it is coupled with a semantic building twin. For bathroom-centered indoor safety analysis, posture and motion descriptors should not be interpreted in isolation, but in relation to bathroom objects and functional zones. This requirement motivates the proposed human-building DT framework, in which the human twin provides dynamic behavior representation, while the semantic bathroom twin provides spatial and contextual grounding. 

\subsection{HBI digital twins}
Human-building interaction (HBI) frames buildings as interactive environments that shape occupant behavior, rather than as passive containers of activity~\cite{taherkhani2023human,kim2023human}. This perspective is highly relevant to bathroom safety, where movement occurs in close relation to fixtures, wet regions, room boundaries, and supportable or collision-prone surfaces. Safety-relevant events are therefore not only motion events, but also body-environment interaction processes.

Existing DT and sensing studies have not yet fully operationalized this perspective for event-level indoor safety analysis. Building-oriented DTs often emphasize geometry, assets, systems, or environmental variables, while human-centered sensing systems mainly classify activities or abnormal events. In both cases, the relation between human motion and semantic environmental context is often weakly represented. This limits the interpretation of events such as support-seeking behavior, contact with fixtures, or impact location~\cite{shishov2024interactions,valipoor2020falls}. Accordingly, the proposed framework treats HBI as a computational problem within a DT environment. By coupling a skeleton-based human twin with a semantic bathroom twin, the framework enables human motion to be interpreted relative to objects, functional zones, and safety-relevant regions. This supports a shift from isolated motion recognition toward spatially grounded and interaction-aware safety interpretation.

\subsection{Bathroom safety and the built-environment context}
Bathrooms are safety-sensitive indoor micro-environments because they combine local wetness, fixture proximity, limited space, and frequent posture transitions within a compact spatial setting~\cite{ng2022bathroom, penner2013bathroom}. Daily bathroom activities such as washing, turning, and stepping across wet areas often occur close to walls, toilets, floors, and other supportable or collision-prone surfaces. Therefore, bathroom safety is not only a matter of individual behavior or environmental hazards, but a coupled issue shaped by how human movement unfolds in relation to built-environment conditions~\cite{lopez2012built, pinter2018impact}.

Existing bathroom safety studies have provided important evidence from design, accessibility, and intervention perspectives. This literature has identified the relevance of fixture arrangement, support elements, spatial accessibility, and bathroom modifications for reducing motion-related risks and improving safety~\cite{afifi2015geriatric,valipoor2020falls,ng2022bathroom}. These studies establish the bathroom as an important built-environment context for indoor safety research. However, most of them remain descriptive, design-oriented, or intervention-oriented. They usually identify risk factors or propose environmental modifications, but they do not provide a framework for dynamically representing how human posture, movement, object proximity, and body-environment interaction jointly contribute to safety-relevant events.

\subsection{Research gap and positioning of the present study}
The reviewed literature indicates that built-environment digital twins, human-centered sensing, human-building interaction, and bathroom safety research have developed largely as parallel research streams. Building DT studies provide important foundations for spatial representation, semantic modeling, and data-connected virtual environments, but they remain mainly building-facing and are rarely developed around the safety needs of older adults in specific domestic micro-environments. Human-centered sensing and pose-based motion analysis, in contrast, provide dynamic representations of posture, movement, and abnormal events, including fall-related behaviors, but they often treat the surrounding environment as background context. This limitation is particularly important for older adult bathroom safety because safety-relevant events in bathrooms are not determined by body motion alone. They are shaped by the interaction between human vulnerability, activity demands, and local environmental conditions such as wet surfaces, limited clearance, fixture proximity, level changes, and insufficient support.

HBI research highlights the importance of relational analysis between occupants and buildings, while bathroom safety studies identify the environmental relevance of fixtures, wet areas, support elements, and spatial constraints. However, these perspectives have not yet been sufficiently integrated into a DT framework for event-level safety analysis in older adult bathroom use. Existing bathroom safety studies can identify environmental hazards or recommend modifications, but they usually do not provide a dynamic representation of how an older person moves, transfers, seeks support, approaches fixtures, or contacts the surrounding environment during daily bathroom activities. Conversely, existing fall-detection and human-sensing systems can recognize motion states or abnormal events, but they often lack semantic knowledge of the bathroom environment and therefore provide limited explanation of where, how, and why a safety-relevant event occurs. This leaves a clear gap at the intersection of older adult safety, bathroom-centered built-environment analysis, human motion representation, and interaction-aware DT modeling. Current research still lacks a bathroom-centered human-building digital twin framework that can interpret safety-relevant events for older adults through explicit body-environment relations.

The present study is positioned to address this gap by proposing a bathroom-centered human-building DT framework for interaction-aware indoor safety analysis with a specific emphasis on older adult bathroom safety. The framework combines a LiDAR-informed semantic bathroom twin, an mmWave-informed skeleton-based human twin, spatial-semantic coupling, and interaction-aware analytics within a unified digital environment. Rather than treating the bathroom as a passive scene or the older occupant as an isolated motion source, the proposed framework models safety-relevant events as spatially grounded human-environment interaction processes. In this way, the study aims to move beyond passive visualization, generic bathroom hazard description, and isolated fall recognition toward a more interpretable, interaction-aware, and semantically grounded approach to analyzing older adults' daily bathroom safety. To further clarify the positioning of the present study, Table~\ref{tab:related_work_comparison} summarizes representative studies across built-environment DTs, human-centered sensing, HBI, and bathroom safety research. The comparison focuses on whether each study provides semantic environmental representation, dynamic human representation, explicit body-environment coupling, bathroom-specific safety analysis, and an older-adult safety orientation. This synthesis is not intended to provide an exhaustive review of all related work, but to highlight the methodological gap that motivates the proposed framework.

\begin{table*}[htbp]
\centering
\caption{Comparison of representative studies and the proposed framework}
\label{tab:related_work_comparison}
\footnotesize
\setlength{\tabcolsep}{4pt}
\renewcommand{\arraystretch}{1.25}
\begin{tabularx}{\textwidth}{p{1.5cm}p{2cm}p{1.2cm}p{1.2cm}p{1.4cm}p{1.5cm}p{2cm}p{1.5cm}p{3.7cm}}
\toprule
\textbf{Study} & \textbf{Main domain} & \textbf{Building semantics} & \textbf{Behavior modeling} & \textbf{HBI} & \textbf{Bathroom focus} & \textbf{Environmental context} & \textbf{Real-data} & \textbf{Main limitation relative to this study} \\
\midrule

AlBalkhy et al.(2024)~\cite{albalkhy2024digital}
& Built-environment digital twin review 
& Yes 
& No 
& No 
& No 
& Partial 
& Partial 
& Focuses on DT definition, structure, and building-side applications; does not provide a bathroom-centered, behavior-aware interaction framework. \\

Taherkhani et al.(2023) ~\cite{taherkhani2023human}
& HBI bibliometric review 
& Partial 
& Partial 
& Partial 
& No 
& Partial 
& No 
& Clarifies HBI as an interdisciplinary field, but does not implement event-level safety analytics or a digital twin platform for bathrooms. \\

Kim et al.(2023) ~\cite{kim2023human}
& HBI for indoor environmental control 
& Partial 
& Yes 
& No 
& No 
& Yes 
& Partial 
& Emphasizes occupant-building interaction for control and comfort, not body-part-to-object interaction or bathroom safety events. \\

Sykes et al.(2025) ~\cite{sykes2025next}
& Pose-based fall detection review 
& No 
& Yes 
& No 
& No 
& No 
& Partial 
& Shows the value of skeletal representations for fall monitoring, but treats the environment largely as background rather than a semantic interaction space. \\

Li et al.(2025) ~\cite{li2025decade}
& Wearable fall detection review 
& No 
& Yes 
& No 
& No 
& No 
& Yes 
& Strong on person-centric sensing, but weak on spatial semantics, object interaction, and room-level environmental interpretation. \\

Shishov et al.(2024) ~\cite{shishov2024interactions}
& Real-life fall interaction analysis 
& No 
& Yes 
& Yes 
& No 
& No 
& Yes 
& Demonstrates that falls often involve environmental objects beyond the floor, but does not provide a digital twin framework or bathroom-based spatial analytics. \\

Afifi et al.(2015) ~\cite{afifi2015geriatric}
& Geriatric bathroom design review 
& Yes 
& No 
& No 
& Yes 
& Partial 
& No 
& Strongly relevant to bathroom safety design, but design-oriented rather than computationally integrative; no human twin or event logging. \\

Valipoor et al.(2020) ~\cite{valipoor2020falls}
& Built-environment fall risk review 
& Yes 
& No 
& No 
& Partial 
& Partial 
& No 
& Identifies interior-scale risk elements, but does not operationalize them in a behavior-aware DT platform. \\

Ng et al.(2022) ~\cite{ng2022bathroom}
& Bathroom modifications among older adults 
& Yes 
& No 
& No 
& Yes 
& No 
& Yes 
& Highlights the practical importance of bathroom modifications after falls, but does not model behavior, interaction sequences, or digital-twin-based risk analysis. \\

\textbf{This study} 
& \textbf{Bathroom-centered human-building digital twin} 
& \textbf{Yes} 
& \textbf{Yes} 
& \textbf{Yes} 
& \textbf{Yes} 
& \textbf{Yes} 
& \textbf{No} 
& \textbf{Integrates semantic bathroom representation, skeleton-based behavior, object-level interaction logging, preliminary risk indicators, and a planned pathway for future in-situ data updating.} \\

\bottomrule
\end{tabularx}
\end{table*}

\begin{figure*}[t]
    \centering
    \includegraphics[width=18cm, trim=3 3 3 3,
        clip]{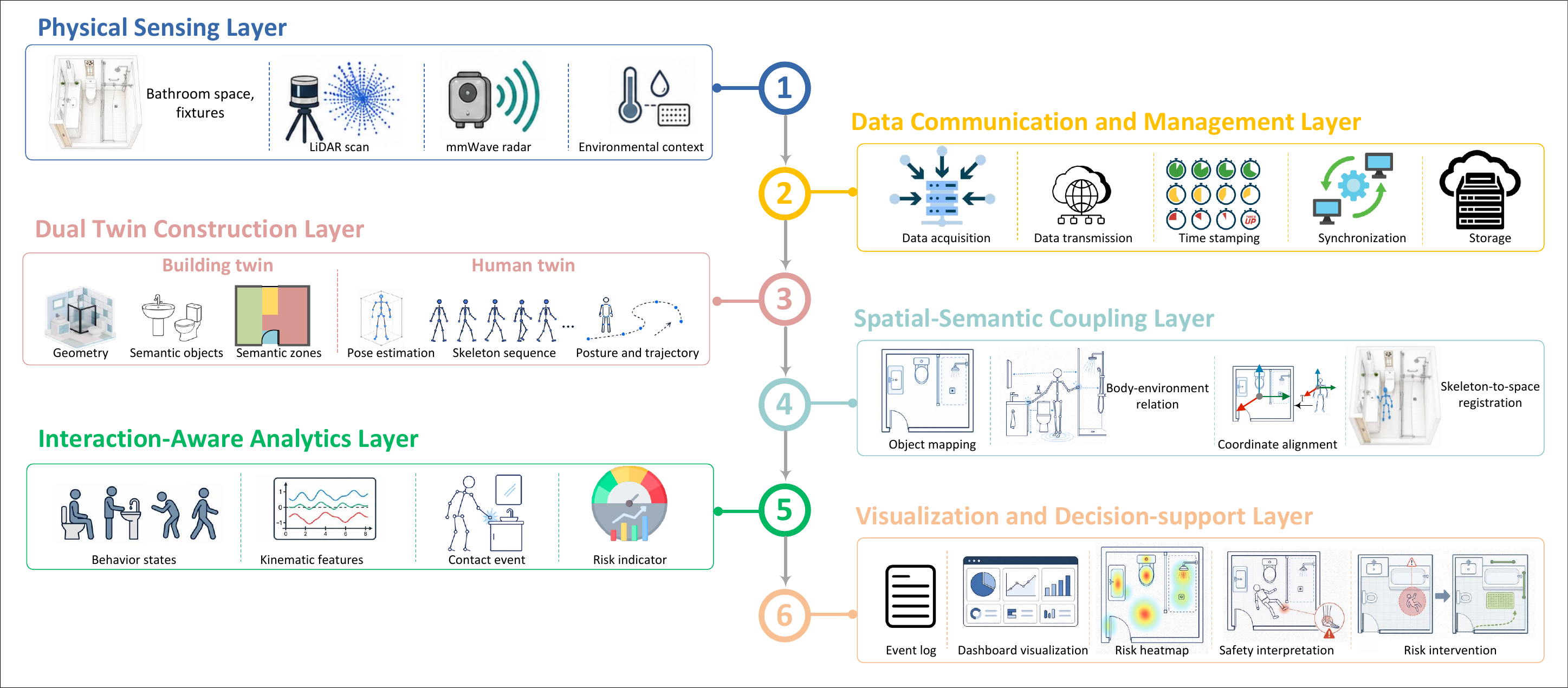}
    \caption{Six-layer architecture of the proposed bathroom-centered human-building digital twin for interaction-aware indoor safety analysis.}
\label{fig:six_layer_architecture}
\end{figure*}

\begin{figure*}[t]
    \centering
    \includegraphics[width=\linewidth, trim=3 3 3 3,
        clip]{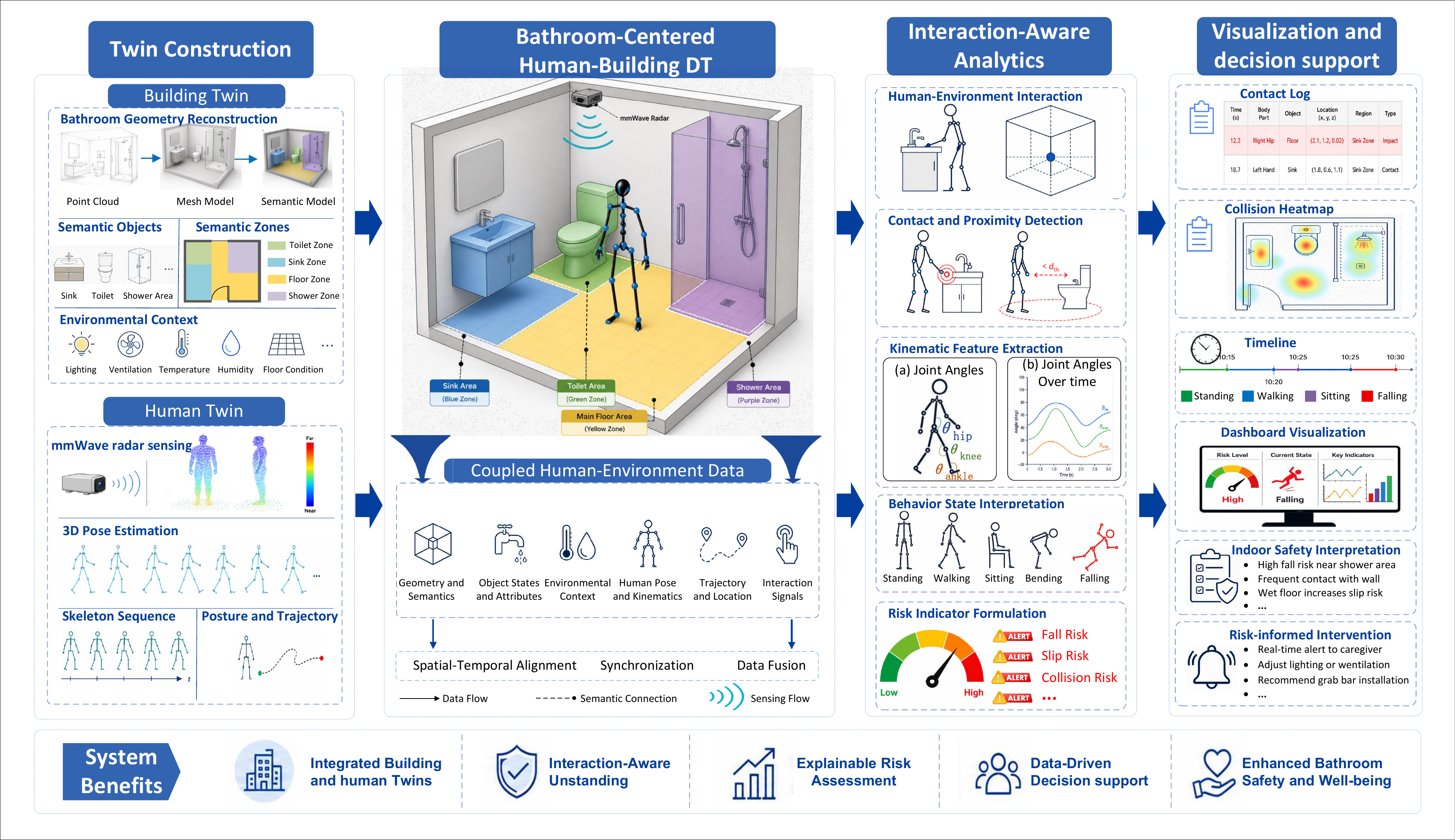}
    \caption{Overall functional workflow of the proposed bathroom-centered human-building digital twin, from twin construction and human-environment coupling to interaction-aware analytics and decision-support outputs.}
\label{fig:overall_functional_framework}
\end{figure*}

\begin{figure}[t]
    \centering
    \includegraphics[width=\linewidth,trim=3 3 3 3,
        clip]{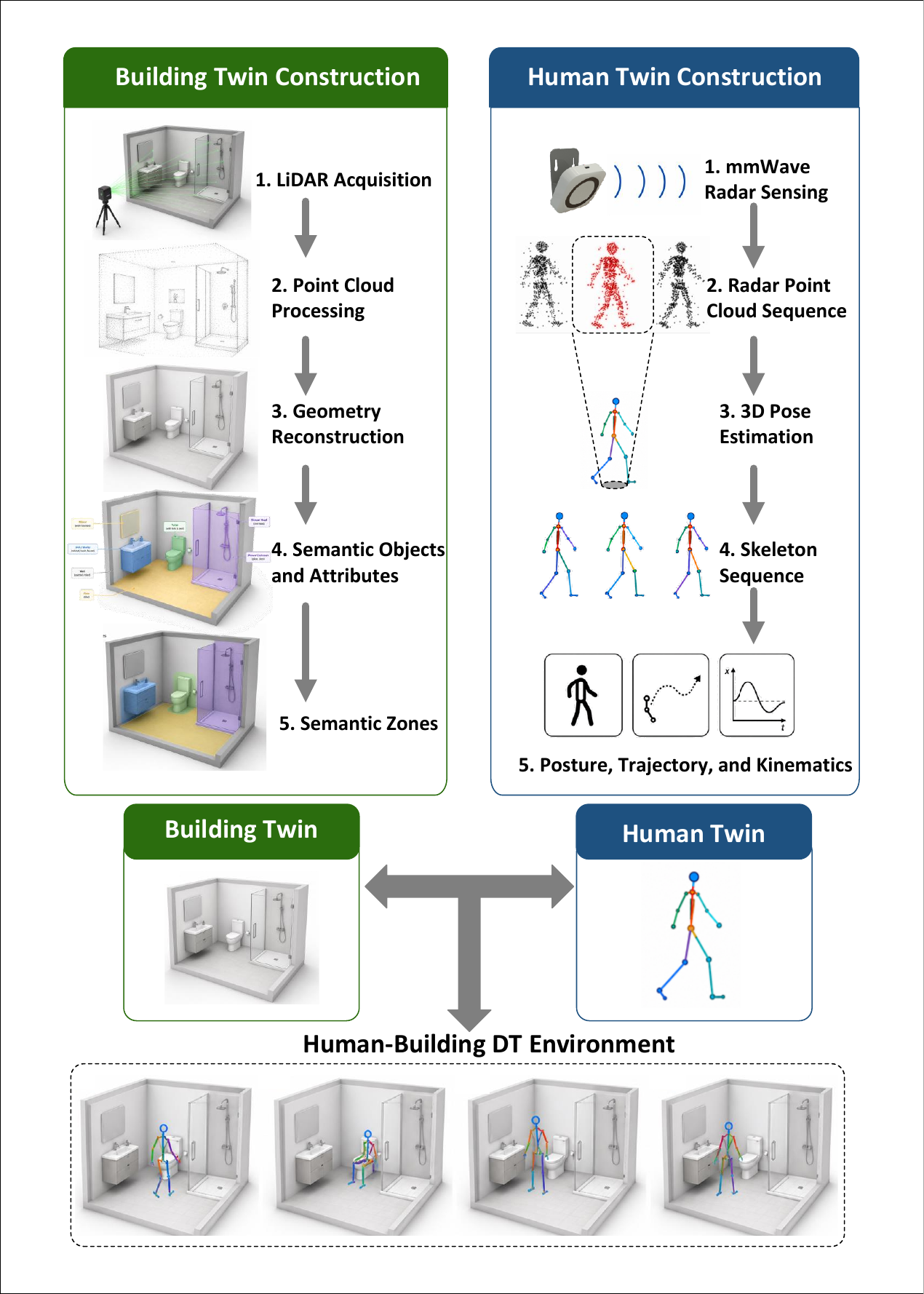}
    \caption{Dual-twin construction workflow, including semantic bathroom twin construction and skeleton-based human twin construction.}
\label{fig:twin_construction_workflow}
\end{figure}

\section{Description of the proposed DT architecture}
\subsection{Architecture Overview}
The overview of the proposed framework is presented in Fig.~\ref{fig:six_layer_architecture}. This study proposes a bathroom-centered human-building digital twin (DT) architecture for interaction-aware indoor safety analysis. Unlike conventional building-operation DTs that mainly focus on equipment monitoring, energy management, or facility control, the proposed architecture is designed to represent and interpret the coupled relationship between human motion-related safety and bathroom spatial semantics. The bathroom is therefore not treated as a passive geometric background, but as a semantically structured and safety-relevant indoor micro-environment in which human posture, movement, object proximity, and body-environment interaction jointly shape safety-related events.

The proposed architecture is organized into six layers: the physical sensing layer, data communication and management layer, human-building twin construction layer, spatial-semantic coupling layer, interaction-aware analytics layer, and visualization and decision-support layer. Table~\ref{tab:six_layer_architecture} summarizes the six layers, their major components, and their roles in the proposed workflow. These layers form a sensing-to-twin-to-analysis workflow. Physical and sensing data are first acquired from the bathroom environment and occupant motion. The acquired data are then transmitted, synchronized, and transformed into structured representations for twin construction. The building twin and human twin are subsequently integrated into a shared coordinate system, enabling body motion to be interpreted relative to bathroom objects, functional zones, and safety-relevant events. Based on this coupled representation, the framework supports behavior state abstraction, kinematic feature extraction, interaction event representation, risk interpretation, and visual decision support. The overall functional workflow is further illustrated in Fig.~\ref{fig:overall_functional_framework}, which complements the six-layer architecture by showing how twin construction, spatial--semantic coupling, interaction-aware analytics, and decision-support outputs are connected as an operational sequence.

\begin{table*}[htbp]
\centering
\caption{Six-layer architecture of the proposed bathroom-centered human-building digital twin}
\label{tab:six_layer_architecture}
\footnotesize
\setlength{\tabcolsep}{5pt}
\renewcommand{\arraystretch}{1.25}
\begin{tabularx}{\textwidth}{p{0.8cm}p{2.0cm}p{4.7cm}X}
\toprule
\textbf{Layer} & \textbf{Name} & \textbf{Main components} & \textbf{Role in the framework} \\
\midrule
1 & Physical sensing layer & Bathroom space, fixtures, LiDAR scan, mmWave radar, optional environmental sensors & Captures the physical bathroom environment and occupant-related sensing inputs. \\
2 & Data communication and management layer & Data acquisition, transmission, timestamping, synchronization, storage, interface between sensing and computation modules & Converts heterogeneous sensing outputs into structured data streams for twin construction and analysis. \\
3 & Human-building twin construction layer & Semantic building twin, skeleton-based human twin, object and zone definitions & Builds complementary digital representations of the bathroom environment and human motion. \\
4 & Spatial-semantic coupling layer & Coordinate alignment, skeleton-to-space registration, object/zone mapping, body-environment relation modeling & Places the human twin inside the semantic bathroom twin so that motion can be interpreted relative to local spatial context. \\
5 & Interaction-aware analytics layer & Behavior states, kinematic descriptors, contact/proximity events, risk indicators & Extracts safety-relevant information from coupled human-environment representations. \\
6 & Visualization and decision-support layer & Unity-based DT scene, skeleton animation, event log, contact records, heatmap, dashboard & Communicates analytical outputs for event review, safety interpretation, and future design or intervention support. \\
\bottomrule
\end{tabularx}
\end{table*}

The six-layer structure is intended to clarify the role of each module and to avoid reducing the proposed DT to a simple three-dimensional visualization prototype. At the current stage, the framework should be understood as a proof-of-concept architecture rather than a fully validated deployment system. Some modules, such as semantic bathroom reconstruction, skeleton reconstruction and visualization, contact-related event representation, and risk indicators, are implemented in the Unity-based prototype, whereas richer information, like environmental sensing, continuous in-situ data updating, remains planned extensions.

\subsection{Physical sensing layer}
The physical sensing layer defines the physical entities and sensing sources that are represented in the DT. In the proposed framework, the bathroom is treated as a safety-critical indoor micro-environment composed of room boundaries, floor surfaces, walls, fixtures, and functional activity regions. Typical safety-relevant bathroom elements include the toilet, sink, shower or bathing area, floor, wall, and other supportable or collision-prone surfaces. These elements are relevant not only because of their geometry, but also because they provide functional and safety-related meanings. For example, a floor surface may be associated with slip or impact risk, and a sink or toilet may become involved in support-seeking or collision events.

Two sensing streams are considered in the current framework. The first stream is an environmental sensing stream for constructing the building twin. LiDAR scanning is used to acquire a point cloud representation of the bathroom environment. This point cloud provides the geometric basis for reconstructing the bathroom layout, including the dimensions and spatial arrangement of major fixtures and room boundaries. The second stream is a human sensing stream for constructing the human twin. mmWave radar is used to sense human motion in a privacy-aware manner. The radar stream provides motion-related point clouds that are processed by a 3D human pose estimation model to derive a skeleton-based representation of the occupant. Compared with direct video capture, this sensing pathway is better aligned with privacy-sensitive bathroom scenarios while still supporting dynamic motion interpretation. The architecture also allows environmental sensing channels, such as temperature, humidity, and carbon dioxide concentration. These channels are not treated as mandatory components of the current prototype, but they provide an extension pathway for future real-world deployment. 

\subsection{Data communication and management layer}
The data communication and management layer connects physical sensing with computational modules. Its function is to acquire, transmit, synchronize, and organize heterogeneous sensing data before these data are used for twin construction and analytical processing. 

In the proposed framework, LiDAR-derived environmental data are typically static, whereas mmWave-derived human motion data are dynamic. The LiDAR-derived environmental stream supports point-cloud-guided reconstruction of the bathroom space. The mmWave-derived human stream supports pose inference and skeleton-sequence generation. Between sensing and twin generation, the raw sensing outputs are transmitted to the corresponding computational modules for preprocessing. For the human sensing stream, this includes radar point-cloud preprocessing, frame organization, and preparation for 3D pose estimation. For the environmental stream, this includes point-cloud cleaning, geometric checking, and preparation for semantic reconstruction in the DT environment. In the current proof-of-concept prototype, the data pipeline is implemented at a modular level to support data import, skeleton playback, and prototype-level analysis in Unity. In future in-situ deployment, this layer can be extended toward real-time streaming, local or cloud storage, multi-sensor synchronization, and continuous updating of the DT using data collected during actual bathroom use.

\subsection{Dual twin construction layer}
As illustrated in Fig.~\ref{fig:twin_construction_workflow}, the human-building twin construction layer builds two complementary digital representations: a semantic building twin of the bathroom and a skeleton-based human twin of the occupant. This layer is central to the proposed framework because bathroom safety is represented as a coupled human-environment problem rather than as a purely spatial or purely motion-based problem.

\subsubsection{Building twin construction}
The building twin serves as the spatial and semantic foundation of the proposed DT. Let the LiDAR point cloud of the bathroom environment be denoted by:
\begin{equation}
\mathcal{P}_b = \{\mathbf{e}_1,\mathbf{e}_2,\dots,\mathbf{e}_n\}, \quad \mathbf{e}_i \in \mathbb{R}^3 .
\end{equation}
This point cloud captures the geometry of the indoor environment, including room boundaries, floor surfaces, fixture dimensions, and the spatial extent of safety-relevant objects. In the present framework, the LiDAR-derived point cloud provides the basis for reconstructing the bathroom environment in the DT platform. The resulting model preserves not only the overall bathroom geometry, but also object size, object arrangement, and layout-sensitive relations that are important for safety analysis.

The reconstructed bathroom is further organized semantically. Major environmental components such as the floor, wall, sink, toilet, and bathing region can be assigned object identities and semantic roles. The set of semantic objects can be expressed as:
\begin{equation}
\mathcal{O}=\{o_1,o_2,\dots,o_k\},
\end{equation}
where each object $o$ is associated with geometric attributes, functional identity, and safety-relevant properties. In addition to object-level semantics, the bathroom can be partitioned into functional or safety-related zones:
\begin{equation}
\mathcal{L}=\{\ell_1,\ell_2,\dots,\ell_q\},
\end{equation}
where each semantic zone may correspond to an entrance zone, sink-use zone, toilet-use zone, or wet-risk zone. This semantic organization allows later analytical modules to interpret human motion not only by coordinates, but also by object and zone context.

\subsubsection{Human twin construction}
The human twin provides a dynamic representation of the occupant in the bathroom environment. Let the mmWave-radar-derived point cloud sequence at frame $t$ be denoted by:
\begin{equation}
\mathcal{P}_h^t=\{\mathbf{r}_1^t,\mathbf{r}_2^t,\dots,\mathbf{r}_m^t\}, \quad \mathbf{r}_m^t\in\mathbb{R}^3 ,
\end{equation}
where $m$ is the number of radar points at frame $t$. After acquisition, transmission, and preprocessing, the radar-derived data are fed into the 3D human pose estimation model developed in our previous work~\cite{su2025posegraphnet}. The model outputs a time-varying skeleton representation:
\begin{equation}
\mathcal{J}^t=\{\mathbf{j}_1^t,\mathbf{j}_2^t,\dots,\mathbf{j}_k^t\}, \quad \mathbf{j}_k^t\in\mathbb{R}^3,
\end{equation}
where $\mathbf{j}_k^t$ denotes the estimated position of joint $k$ at frame $t$. The skeleton sequence over an observation period can then be written as:
\begin{equation}
\mathcal{K}_{1:T}=\{\mathcal{J}^1,\mathcal{J}^2,\dots,\mathcal{J}^T\}.
\end{equation}
This skeleton sequence forms the human twin in the proposed framework. Compared with directly using raw radar data for downstream interpretation, the skeleton representation provides a compact, structured, and interpretable abstraction of posture and motion. It also facilitates later processing such as behavior state understanding, kinematic feature extraction, body-part-level relation analysis, and event interpretation.

It should be added that the human twin in this study should not be understood as a full physiological or individualized replica of a person. It is a behavior-aware digital representation derived from sensing and pose inference. Its primary role is to provide a dynamic human component that can be spatially coupled with the semantic bathroom twin for indoor safety analysis.

\subsection{Spatial-semantic coupling layer}
The spatial-semantic coupling layer integrates the building twin and human twin into a shared analytical space. This layer is a key distinction between the proposed framework and conventional fall-detection or activity-recognition systems. A motion label, such as walking, sitting, or falling, is insufficient for bathroom safety analysis unless it is interpreted relative to where the event occurs, which object is nearby, which body part is involved, and whether the event overlaps with a safety-relevant region. Therefore, the purpose of this layer is to transform skeleton-based motion into spatially and semantically contextualized human-environment relations.

\subsubsection{Twin coordinate alignment}
To integrate the two twins in a common analytical space, the framework defines a bathroom-centered coordinate system. Let the bathroom-centered origin be denoted by:
\begin{equation}
\mathbf{c}=(x_c,y_c,z_c),
\end{equation}
and let the estimated position of joint $k$ in the human sensing frame at frame $t$ be denoted by $\mathbf{j}_k^t$. Since the human twin and building twin are derived from different sensing sources, a spatial transformation is required to map the human skeleton into the bathroom coordinate system. A general rigid transformation can be written as:
\begin{equation}
\hat{\mathbf{j}}_k^t=\mathbf{R}\mathbf{j}_k^t+\mathbf{d},
\end{equation}
where $\hat{\mathbf{j}}_k^t$ is the aligned joint position in the bathroom coordinate system, $\mathbf{R}$ is a rotation matrix, and $\mathbf{d}$ is a translation vector. Depending on the implementation, this transformation can be obtained through calibration, manual registration, or known spatial relationships between the radar sensor and the bathroom model.

\subsubsection{Semantic mapping of body-environment relations}
After coordinate alignment, the skeleton can be mapped to semantic objects and zones. For each aligned joint or body segment, the system can compute its relationship to environmental objects and regions. For example, the distance between a joint $\hat{\mathbf{j}_i^t}$ and an environmental object $o_k$ at frame $t$ can be generally expressed as:
\begin{equation}
d_{i,k}(t)=\min_{\mathbf{x}\in o_k}
\left\|\hat{\mathbf{j}}_i^t-\mathbf{x}\right\|_2.
\end{equation}
This formulation supports proximity analysis, potential contact detection, and object-level relevance assessment. Similarly, the semantic zone occupied by a joint or body segment can be identified through a zone assignment function:
\begin{equation}
\ell(t)=\Phi(\hat{\mathbf{j}}_k^t,\mathcal{L}),
\end{equation}
where $\Phi(\cdot)$ maps the aligned body coordinate to a semantic bathroom zone. Through this process, low-level skeleton trajectories are transformed into interpretable body-environment relations, such as movement near the sink, falling within a wet-risk zone, or support-seeking contact with a fixture. This layer therefore provides the representational basis for interaction-aware analytics.

\subsection{Interaction-aware analytics layer}
The interaction-aware analytics layer extracts safety-relevant information from the coupled human-building representation. It is responsible for converting aligned skeleton sequences and semantic bathroom context into behavior states, kinematic descriptors, interaction event records, and risk indicators. This layer shifts the analytical focus from simple event recognition toward event explanation.

\subsubsection{Behavior state representation}
Based on the human skeleton obtained from the human-twin construction module, the continuous motion sequence can be further interpreted into discrete behavior states using AI-based skeleton action recognition methods. Specifically, the 3D joint positions, body posture, and temporal motion patterns of the skeleton sequence can be used as inputs for behavior recognition models~\cite{shi2019two,duan2022revisiting}. Through this process, the DT can infer the occupant's current behavior state, such as falling. Let the behavior state at frame $t$ be denoted by $s_t$, and the state sequence over an observation period be written as:
\begin{equation}
\mathcal{S}_{1:T}=\{s_1,s_2,\dots,s_T\}.
\end{equation}

This behavior state representation serves as an intermediate analytical layer between low-level skeleton trajectories and higher-level safety interpretation. It allows subsequent modules to analyze not only the motion itself, but also how the inferred behavior state relates to bathroom objects, functional zones, and safety-relevant interaction events.

\subsubsection{Kinematic feature extraction}
Based on the aligned skeleton sequence, the framework extracts kinematic descriptors that characterize body motion and posture evolution. The hip joint can be used as a reference point for characterizing global body motion. Let the aligned hip position be denoted by $\hat{\mathbf{j}}_h^t$. The corresponding hip velocity is defined as:
\begin{equation}
\mathbf{v}_h(t)=\frac{\hat{\mathbf{j}}_h^t-\hat{\mathbf{j}}_h^{t-\Delta t}}{\Delta t},
\end{equation}
where $\Delta t$ is the temporal sampling interval. The scalar movement speed is then:
\begin{equation}
v_h(t)=\|\mathbf{v}_h(t)\|_2 .
\end{equation}
To represent motion extent over a time interval, cumulative displacement can be defined as:
\begin{equation}
D_h(t_a,t_b)=\sum_{t=t_a+\Delta t}^{t_b}\|\hat{\mathbf{j}}_h^t-\hat{\mathbf{j}}_h^{t-\Delta t}\|_2 .
\end{equation}
Posture-related features can also be extracted from the skeleton configuration. Let $\mathbf{u}(t)$ and $\mathbf{w}(t)$ denote two body-segment vectors. An angle between two body-segment vectors can be written as:
\begin{equation}
\theta(t)=\cos^{-1}\left(\frac{\mathbf{u}(t)\cdot\mathbf{w}(t)}{\|\mathbf{u}(t)\|_2\|\mathbf{w}(t)\|_2}\right).
\end{equation}
Depending on the implementation, such measures can describe trunk inclination, hip flexion, limb orientation, or abrupt posture change. These features provide quantitative descriptors for interpreting unstable movement, fall-related motion, and transition-related safety events.

\subsubsection{Interaction event representation}
A key objective of the framework is to represent human motion as body-environment interaction rather than isolated movement. The system can analyze the spatial relationship between body parts and bathroom objects over time. A body-environment interaction event is defined as:
\begin{equation}
e_n=\left(t_n,j_i,o_k,\mathbf{c}_n,\ell_n,\alpha_n\right),
\end{equation}
where $t_n$ is the event time, $j_i$ is the involved body part, $o_k$ is the involved environmental object, $\mathbf{c}_n$ is the event location, $\ell_n$ is the corresponding semantic region, and $\alpha_n$ is the interaction type. Additionally, $\alpha_n$ can be denoted as direct contact, proximity-based hazard relation, support-related relation, impact-related relation, or repeated contact. This representation allows the DT to answer questions that pure motion recognition cannot address, such as where a fall occurs, which object becomes involved, which body part contacts the environment first, and how the event unfolds relative to local bathroom layout. Table~\ref{tab:contact_event_log} provides a simplified example of how such interaction events are structured in the prototype. Rather than recording contact as an isolated collision, each event is linked to time, body part, object, semantic zone, interaction type, and safety relevance.

\begin{figure*}[t]
    \centering
    \includegraphics[width=17 cm]{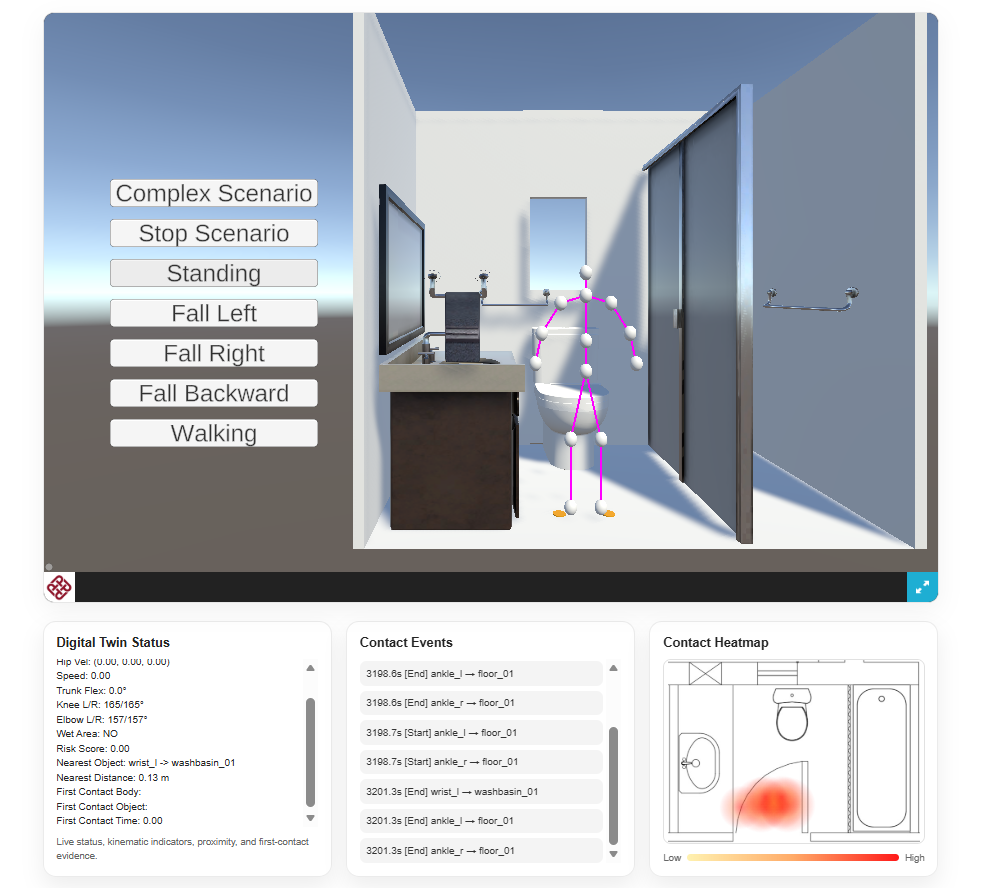}
    \caption{Unity-based proof-of-concept prototype of the bathroom-centered human–building digital twin.}
\label{fig:Unity_based_proof_of_concept}
\end{figure*}

\begin{figure*}[t]
    \centering
    \includegraphics[width=\linewidth,trim=3 3 3 3,
        clip]{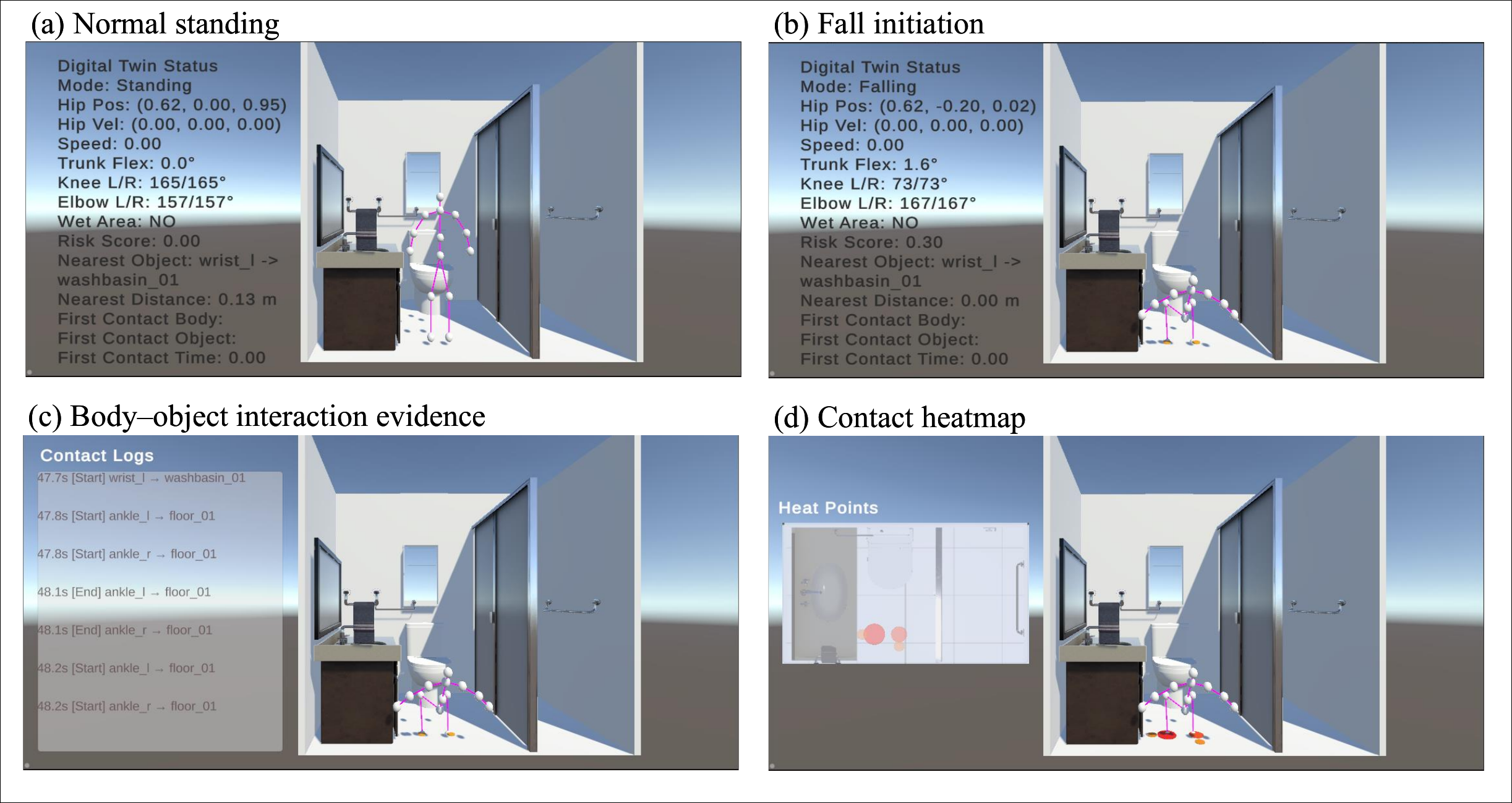}
    \caption{Example prototype scenario showing interaction-aware event interpretation.}
\label{fig:Example_prototype_scenario}
\end{figure*}

\subsubsection{Risk indicator formulation}
Indoor safety risk is represented in the framework as a composite indicator derived from behavior state, kinematics, interaction, and environmental context. At the current stage, this formulation is intended as an interpretable prototype-level measure rather than a clinically validated predictive score. Let the time-varying risk indicator be denoted by:
\begin{equation}
R(t)=w_sR_s(t)+w_kR_k(t)+w_iR_i(t)+w_eR_e(t),
\end{equation}
where $R_s(t)$ is the state-related term, $R_k(t)$ is the kinematic term, $R_i(t)$ is the interaction-related term, $R_e(t)$ is the environmental-context term, and $w_s$, $w_k$, $w_i$, and $w_e$ are nonnegative weighting parameters.

The state-related term reflects whether the current motion belongs to a relatively stable or unstable category. The kinematic term captures rapid motion, abrupt displacement, marked posture change, or fall-related dynamics. The interaction-related term reflects the occurrence and severity of body-environment interaction, especially when semantically safety-critical objects or surfaces are involved. The environmental-context term captures local contextual modifiers such as wet-risk zones. For an observation interval, an event-level risk score can be defined as:
\begin{equation}
R_{\mathrm{event}}=\frac{1}{N}\sum_{m=1}^{N}R(t_m),
\end{equation}
where $t_m$ denotes the sampled time points within the interval. This aggregate measure provides a compact summary of safety-relevant conditions over time and supports later comparative interpretation.

\begin{table*}[htbp]
\centering
\caption{Example of structured contact-event log used in the prototype}
\label{tab:contact_event_log}
\begin{tabular}{ccccccc}
\hline
Event ID & Time (s) & Body part & Object & Semantic zone & Interaction type & Safety relevance \\
\hline
E1 & 3.42 & Right hand & Sink edge & Sink zone & Proximity & Possible support-seeking behavior \\
E2 & 5.18 & Left foot & Floor & Wet-risk zone & Contact & Movement through slippery area \\
E3 & 6.74 & Hip & Floor & Shower zone & Impact contact & Potential fall-related impact \\
E4 & 7.06 & Left arm & Wall & Shower zone & Contact & Secondary contact after instability \\
E5 & 8.20 & Right hand & Floor & Shower zone & Impact contact & Possible post-fall inactivity \\
\hline
\end{tabular}
\end{table*}

\subsection{Visualization and decision-support layer}
The visualization and decision-support layer provides the user-facing representation of the proposed DT. The current prototype is implemented in Unity, which serves as the integration platform in which the LiDAR-informed building twin and the mmWave-derived human twin are visualized, aligned, and analyzed in the same digital space. In this implementation, Unity is not used merely as a rendering tool, but as the operational environment in which the sensing-to-twin-to-analysis pipeline is demonstrated.

The prototype supports the display of the reconstructed bathroom environment, the aligned 3D human skeleton, motion-related descriptors, and interaction-aware outputs. It can also support event logs, contact records, collision heatmaps, behavior timelines, and risk-related dashboard elements. These outputs are intended to assist post-event review and safety interpretation by making event processes visible and spatially contextualized. For example, instead of only indicating that a fall-related event has occurred, the DT can be designed to show the semantic region of the event, the involved object or surface, the body part associated with contact, and the local motion pattern before and after the event. 

The decision-support function of this layer should be interpreted cautiously at the current stage. The prototype does not yet provide a fully validated clinical risk assessment or an autonomous intervention system. Rather, it provides a methodological and computational basis for future bathroom safety analysis, layout-sensitive design evaluation, caregiver-oriented event review, and real-world deployment studies. Through this layer, the proposed DT translates coupled human-environment data into interpretable visual and analytical outputs.

\subsection{Prototype implementation and demonstration}
To operationalize the proposed framework, a Unity-based proof-of-concept prototype was developed to demonstrate how the main components of the bathroom-centered human-building digital twin can be integrated within a shared digital environment. The prototype was developed to examine whether the proposed sensing-to-twin-to-analysis workflow can be implemented as an interactive digital environment. The current implementation includes selected modules from the six-layer architecture. On the building side, a reconstructed bathroom scene is represented as a semantic environment containing major fixtures, spatial boundaries, and functional areas. On the human side, skeleton sequences are imported and animated in the Unity environment to form a dynamic human twin. The human skeleton is placed within the bathroom scene so that posture, trajectory, and movement states can be interpreted relative to surrounding objects and zones. In this way, the prototype provides a preliminary implementation of the spatial-semantic coupling layer described above.

Figure~\ref{fig:Unity_based_proof_of_concept} shows the current prototype interface. The main view presents the bathroom scene and the animated human representation. Supporting panels provide digital-twin status information, motion-related indicators, contact-event records, and contact heatmap visualization. These interface components are designed to demonstrate how the prototype translates coupled human-environment data into interpretable outputs for event review and safety-oriented interpretation. Moreover, the prototype also demonstrates how a motion sequence can be transformed into interaction-aware event descriptors. As illustrated in Fig.~\ref{fig:Example_prototype_scenario}, the system can represent a scenario progression from normal standing to fall initiation, body-object interaction evidence, and contact heatmap generation. These descriptors correspond to the event representation defined in Eq.~(16) and provide a structured basis for interpreting where and how a safety-relevant event unfolds within the bathroom layout.

As shown in Table~\ref{tab:prototype_scope}, at the current stage, the prototype implements several core functions: semantic bathroom visualization, skeleton animation, motion-state display, contact-event logging, contact heatmap generation, and contextual risk indication. These functions are sufficient to demonstrate the analytical feasibility of the proposed framework, especially its ability to link occupant motion with bathroom spatial semantics and interaction evidence. However, several components remain prototype-level or planned extensions. Real-time mmWave streaming, continuous multi-sensor synchronization, automatic semantic reconstruction from raw point clouds, validated injury assessment, and autonomous intervention recommendation have not yet been fully implemented. Therefore, the prototype should be interpreted as a feasibility demonstration of the proposed human-building DT workflow rather than as a validated real-world bathroom safety monitoring system.

\begin{table}[htbp]
\centering
\caption{Current implementation scope of the proof-of-concept prototype}
\label{tab:prototype_scope}
\begin{tabular}{p{6cm} c}
\toprule
\textbf{Module} & \textbf{Status} \\
\midrule
Semantic bathroom scene & $\checkmark$ \\
Skeleton visualization & $\checkmark$ \\
Motion-state abstraction & $\triangle$ \\
Contact event log & $\checkmark$ \\
Contact heatmap & $\checkmark$ \\
Risk indicator & $\triangle$ \\
Real-time mmWave streaming & $\circ$ \\
Continuous in-situ updating & $\circ$ \\
Clinically validated injury assessment & $\circ$ \\
Autonomous risk intervention & $\circ$ \\
\midrule
\multicolumn{2}{p{8cm}}{\footnotesize \textit{Note:} $\checkmark =$ implemented in the prototype; $\triangle =$ partially implemented; $\circ =$ not implemented in the current prototype.} 

\end{tabular}
\end{table}

\section{Discussion}
\subsection{Research significance and framework contribution}
The present study should be understood primarily as a framework-oriented contribution to bathroom-centered indoor safety research. Its main significance lies in formalizing bathroom safety as a coupled human-environment analytical problem and operationalizing this perspective in a digital twin setting. In existing research, bathroom safety is often examined through separated perspectives~\cite{afifi2015geriatric,valipoor2020falls,ng2022bathroom,sykes2025next,luo2022pervasive,li2025decade}. Built-environment studies emphasize environmental hazards, accessibility, and design modification, whereas sensing-driven studies emphasize behavior recognition, activity classification, or abnormal event detection~\cite{afifi2015geriatric,valipoor2020falls,ng2022bathroom}. Although both lines of work are valuable, they usually leave unresolved the question of how spatial semantics, occupant motion, and body-environment relations can be jointly represented and interpreted within the same analytical environment. The proposed framework contributes to this gap by organizing bathroom safety analysis around three tightly connected representational layers: a semantic building twin, a behavior-aware human twin, and an interaction-aware analytical layer. In this structure, the bathroom is treated not as a geometric container, but as a semantically meaningful indoor micro-environment composed of objects and regions. The occupant is represented not simply as a moving point or an event label, but as a dynamic skeleton-based human twin whose posture, motion, and state transitions can be interpreted in relation to the surrounding space. Interaction is then elevated from an incidental context to an explicit analytical object through descriptors such as involved body part, environmental object, semantic region, and event location.

A further contribution of the framework is its shift in analytical emphasis from event recognition toward event explanation. In many conventional motion monitoring pipelines, the principal representative is whether an abnormal behavior like a fall occurred or which category a motion sequence belongs to. By contrast, the present framework is intended to describe how a safety-relevant event unfolds in space and time, including where it occurs, which environmental element becomes involved, and how the local bathroom context shapes the event process. This does not eliminate the value of event detection, but it expands the interpretive scope of indoor safety analysis beyond categorical outcomes. From a methodological perspective, the study also contributes an operational sensing-to-twin-to-analysis pipeline. The framework links heterogeneous sensing, twin construction, behavior analysis, interaction-aware interpretation, and risk-oriented visualization. Even at the current proof-of-concept stage, this integration is important because it demonstrates that bathroom safety can be studied as a structured analytical problem. Therefore, the significance of this work lies in establishing a conceptual and operational scaffold for future bathroom-centered digital twin research. Its contribution is therefore better understood as analytical integration and prototype feasibility.

\subsection{Advantages over existing DT approaches}
The first advantage of the proposed framework is that it explicitly couples the building and the occupant as co-interpretable components of the same analytical system. Many existing digital twin implementations in the built environment remain predominantly building-facing: they represent geometry, assets, or environmental variables, but treat the occupant as absent, weakly integrated, or secondary~\cite{albalkhy2024digital,mousavi2024digital,zhang2025review,backlund2024showcasing}. Conversely, many human motion monitoring systems remain predominantly human-facing: they identify motion states or abnormal events, but treat the environment largely as background rather than as a semantically structured participant in risk formation~\cite{sykes2025next,luo2022pervasive,li2025decade,delahoz2014survey,wang2020elderly}. The present framework addresses this separation by integrating a semantic bathroom twin and a skeleton-based human twin within the interpretive workflow.

Moreover, our framework targets the bathroom-centered problem. Bathrooms are compact, semantically dense, and safety-critical indoor micro-environments in which posture transitions, object proximity, support-seeking behavior, and local wetness interact strongly. These characteristics make the bathroom an informative testbed for developing and examining interaction-aware digital twin approaches before extending toward broader domestic environments. In this sense, the specificity of the setting is a strength rather than a limitation at the current stage. These advantages, however, should be interpreted within the scope of the present study. The current work demonstrates analytical integration and prototype feasibility, but it does not yet establish the predictive reliability, deployment robustness, or generalizability required of a mature real-world bathroom safety monitoring system.

\subsection{Implications for bathroom safety analysis and design}
The proposed framework has several implications for how bathroom safety may be studied and interpreted. First, it suggests that bathroom safety should not be analyzed solely through generic fall counts, abstract hazard checklists, or isolated motion labels. Instead, it should be approached as a spatially and behaviorally coupled process in which risk emerges through the relation between occupant movement and environmental configuration. By linking posture transitions, kinematic descriptors, human-environment interaction, and semantic bathroom context, the framework supports a more fine-grained understanding of how safety-relevant situations are formed. Moreover, the framework provides a basis for examining which environmental elements are most relevant to safety in compact bathroom settings. Rather than treating the bathroom as a homogeneous room, the proposed approach distinguishes among fixtures, zones, and contextual regions that may play different roles during instability, support-seeking, impact, or post-event behavior. This makes it possible, at least in principle, to ask more specific questions about bathroom risk, such as which objects are most frequently involved in contact, which zones are associated with repeated interaction, or where event clustering may indicate layout-sensitive vulnerability.

Additionally, the framework points toward a transition from descriptive bathroom safety guidance to analytically informed design exploration. If extended in future work, the digital twin environment could support comparison across alternative layouts, fixture arrangements, support provisions, and wet-area distributions. Such use would be valuable, especially in compact residential contexts, where relatively small spatial modifications may significantly change movement constraints, interaction patterns, and local safety conditions. In this sense, the framework may eventually contribute not only to monitoring-oriented applications but also to layout-sensitive interpretation and design support. At the current stage, these implications should be interpreted as research and methodological value enabled by the framework. Nevertheless, they indicate why a bathroom-centered digital twin may be useful not only for visualization but also for more structured interpretation of indoor safety.

\subsection{Limitations of the current prototype}
Several limitations of the present work should be acknowledged clearly. First, the current implementation remains at the proof-of-concept stage and does not yet constitute a fully validated real-world digital twin deployment. The framework has been operationalized in a prototype environment, but it has not yet been evaluated through long-term in-situ sensing, extended real-world user observations, or real incident datasets.

Specifically, the current human twin is generated by an AI-based pose estimation model from mmWave radar data, rather than being directly measured as a complete human-body replica. As a result, the skeleton representation inevitably contains estimation uncertainty. Potential errors may be introduced by sensing noise, sparse radar point clouds, complex postures, inter-person movement differences, and the limited generalization ability of the model. These errors may propagate to downstream modules, including kinematic feature extraction, behavior-state interpretation, and body-environment interaction analysis. Therefore, the current human twin is suitable for prototype-level feasibility demonstration, but it should not yet be regarded as a fully accurate or individualized representation of real bathroom behavior. Moreover, the current framework does not yet provide comprehensive quantitative validation of event interpretation performance. Although it can generate interpretable outputs such as state transitions, contact-related descriptors, and contextual risk-related variables, these outputs have not yet been systematically benchmarked against annotated bathroom incidents, longitudinal ground truth, or large-scale comparative datasets. Accordingly, the current work should be seen as demonstrating feasibility and analytical expressiveness rather than monitoring accuracy in a validated sense.

Current implementation remains centered on a bathroom-specific setting. This focused scope is methodologically justified for a first prototype, but it also limits direct generalization to multi-room domestic environments, household variability, and broader transitions across everyday residential activities. In addition, the current system does not yet resolve several practical deployment issues that are likely to become central in future work, including long-term sensing robustness, privacy-preserving real-world data capture, synchronization across heterogeneous sensing streams, and stable data association between sensed human behavior and semantic scene elements. These limitations do not negate the value of the framework, but they do define the current boundary of its claims.

\section{Conclusion}
This study proposed a bathroom-centered human-building digital twin framework for interaction-aware indoor safety analysis. The work was motivated by the limitation that bathroom safety is often examined through separated perspectives: built-environment studies usually focus on hazards, accessibility, fixture arrangement, and design modification, whereas sensing-driven studies commonly emphasize activity recognition, fall detection, or abnormal-event classification. Such separated approaches are insufficient for explaining how safety-relevant events are formed in compact bathroom environments, where human posture, movement, object proximity, constrained regions, and built-environment conditions are closely coupled.

To address this gap, the study framed bathroom safety as a dynamic human-environment interaction problem and developed a six-layer digital twin architecture. The proposed architecture consists of a physical sensing layer, data communication and management layer, dual twin construction layer, spatial-semantic coupling layer, interaction-aware analytics layer, and visualization and decision-support layer. Within this structure, the bathroom is represented as a semantic building twin containing objects and safety-relevant regions, while the occupant is represented as a skeleton-based human twin derived from sensing data and AI-based pose estimation. The spatial-semantic coupling layer further aligns the human twin with the bathroom coordinate system, enabling posture, trajectory, kinematic descriptors, contact events, and contextual risk indicators to be interpreted relative to bathroom objects and functional zones. Moreover, a Unity-based proof-of-concept prototype was developed to operationalize components of the proposed framework. The prototype therefore provides evidence of analytical feasibility and demonstrates how the proposed framework can move beyond isolated motion labels toward event-level body-environment interpretation.

The current work should be interpreted primarily as a framework-oriented contribution. It does not yet constitute a fully validated real-world bathroom safety monitoring system. Several components, including real-time mmWave streaming, continuous in-situ updating, automatic semantic reconstruction, environmental sensing integration, validated injury assessment, and autonomous intervention recommendation, remain to be developed and validated. Future work should therefore focus on collecting real-world bathroom data, improving robust human sensing and spatial registration, validating interaction-event descriptors against annotated observations, and examining how bathroom safety indicators can support expert assessment, caregiver review, and design-oriented intervention. Overall, the proposed framework provides a methodological basis for more interpretable, human-centered, and interaction-aware digital twin research in residential indoor safety.

\bibliographystyle{elsarticle-num}
\bibliography{mybibfile}
\end{document}